\documentclass[aps,prapplied,twocolumn,reprint,showpacs,superscriptaddress,footinbib]{revtex4-1}
\usepackage{graphicx} 
\usepackage{dsfont} 
\usepackage{comment}
\usepackage{amsmath} 
\usepackage{graphicx} 
\usepackage{subfigure}
\usepackage[utf8]{inputenc} 
\usepackage{ulem} 
\usepackage{units} 
\usepackage{version} 
\usepackage{verbatim}
\usepackage{color} 
\usepackage{colortbl} 
\usepackage{xcolor}  
\usepackage{fancyref} 
\usepackage[colorlinks=true,citecolor=black,urlcolor=blue,linkcolor=black]{hyperref} 
\usepackage{array} 
\usepackage{booktabs} 
\usepackage{longtable}
\usepackage{tikz}

\newcommand*\circled[1]{\tikz[baseline=(char.base)]{\node[shape=circle,draw,inner sep=1pt] (char) {#1};}}

\begin{document} 

\title{Pumping dynamics of cold-atom experiments in a single vacuum chamber}
\author{Jean-Marc Martin}
\affiliation{SYRTE, Observatoire de Paris, Universit\'e PSL, CNRS, Sorbonne Universit\'e, LNE, 61 avenue 
de l'Observatoire, 75014 Paris, France}

\author{Satyanarayana Bade}
\affiliation{SYRTE, Observatoire de Paris, Universit\'e PSL, CNRS, Sorbonne Universit\'e, LNE, 61 avenue 
de l'Observatoire, 75014 Paris, France}

\author{William Dubosclard}
\affiliation{SYRTE, Observatoire de Paris, Universit\'e PSL, CNRS, Sorbonne Universit\'e, LNE, 61 avenue 
de l'Observatoire, 75014 Paris, France}

\author{Murtaza Ali Khan}
\affiliation{SYRTE, Observatoire de Paris, Universit\'e PSL, CNRS, Sorbonne Universit\'e, LNE, 61 avenue 
de l'Observatoire, 75014 Paris, France}

\author{Seungjin Kim}
\affiliation{SYRTE, Observatoire de Paris, Universit\'e PSL, CNRS, Sorbonne Universit\'e, LNE, 61 avenue 
de l'Observatoire, 75014 Paris, France}

\author{Barry M. Garraway} 
\affiliation{Department of Physics and Astronomy, University of Sussex, Brighton BN1 9QH, UK}

\author{Carlos L. Garrido Alzar} 
\email{carlos.garrido@obspm.fr}
\affiliation{SYRTE, Observatoire de Paris, Universit\'e PSL, CNRS, Sorbonne Universit\'e, LNE, 61 avenue 
de l'Observatoire, 75014 Paris, France} 
\date{\today} 

\begin{abstract} 
A nonlinear analytical model for the pressure dynamics in a vacuum chamber, pumped with a sputter ion pump~(SIP), is 
proposed, discussed and experimentally evaluated. The model describes the physics of the pumping mechanism of SIPs in the 
context of a cold atom experiment. By using this model, we fit pump--down curves of our vacuum system to extract the 
relevant physical parameters characterizing its pressure dynamics. The aim of this investigation is the optimization of 
cold atom experiments in terms of reducing the dead time for quantum sensing using atom interferometry. We 
develop a calibration method to improve the precision in pressure measurements via the ion current in SIPs. Our method is based 
on a careful analysis of the gas conductance and pumping in order to reliably link the pressure readings at the SIP with the 
actual pressure in the vacuum (science) chamber. Our results are in agreement with the existence of essentially two pumping 
regimes determined by the pressure level in the system. In particular we found our results in agreement with the well known 
fact that for a given applied voltage, at low pressures, the discharge current efficiently sputters pumping material from the 
pump's electrodes. This process sets the leading pumping mechanism in this limit. At high pressures, the discharge current 
drops and the pumping is mainly performed by the already sputtered material.
\end{abstract} 

\maketitle

\section{\label{sec:level1}Introduction}
Quantum sensing plays a significant role in the development of the future quantum technologies~\cite{reinhard,leo}. So far, 
this sensing technology has been developed on different physical platforms and, in particular, using cold atoms. Among the 
most relevant realizations one can count atomic microwave and optical clocks~\cite{Norman}, 
magnetometers~\cite{koschorreck,wildermuth} and atom interferometers for inertial sensing~\cite{canuel,dutta,dutta1,barret}, to 
mention a few. In fact, their quantum nature offers a very high sensitivity to measure gravity~\cite{peters,gillot,abend}, 
fundamental constants~\cite{Weiss,bouchendira}, and general relativity effects~\cite{aguilera,fixler,rosi}. Nowadays, these 
experimental realizations have been developed not only to a metrology level, being operated as standards, but also as 
instruments with a maturity that allows industrial and commercial applications~\cite{muquans,aosense}.

A key element to reach the required level of sensitivity to a physical phenomenon is the preparation of a well--controlled state 
of the atoms in terms of their internal and external degrees of freedom. This requires laser cooling in a magneto--optical 
trap (MOT) to sub--Doppler temperatures (on the order of 1 $\mu$K and below), which unavoidably introduces a dead time in the 
measurement process. For atom interferometers, this translates into the well known Dick effect that degrades the stability 
of these devices~\cite{dick,giorgio,quesada,schioppo}. To reduce the MOT loading time, a relatively high background partial 
pressure of the atoms to be cooled~\cite{Monroe} is required, for example $\sim$10$^{-8}$ mbar for $^{87}$Rb atoms. However, 
this high background pressure reduces the available lifetime to perform the desired experiments with the trapped atom 
clouds~\cite{Bjorkholm} and also, it degrades the contrast of the interference fringes leading to a reduction of the 
signal-to-noise ratio of the measurements.

In a typical cold atom experiment the high background pressure problem is overcome, on the one hand, by using two chambers 
connected via a differential pumping stage. In this situation, one chamber (at high pressure) is used as a bright source of 
cold atoms and the other one (at low pressure) as a science chamber. However, this solution is hardly compatible with the 
realization of cold--atom--based compact and miniature sensors. So, on the other hand, when using a single vacuum chamber 
incorporating the atom source (i.e.\ an alkali metal dispenser) after the MOT loading stage the residual background atoms 
need to be pumped out quickly. This is needed in order to preserve a useful level of lifetime of the trapped atoms and to 
avoid an important increase of the dead time. This later situation implies the ability to switch from high 
($\sim$10$^{-8}$ mbar) to low pressure ($\sim$10$^{-11}$ mbar) in a few tenths of ms~\cite{Dugrain}. A very 
promising solution has been recently found~\cite{kitching} which allows one to quickly and reversibly 
control the Rb background pressure in a cell. In this setup, a MOT with up to 10$^6$ atoms has been realized. However, no 
compatibility with a pressure level of $\sim$10$^{-11}$ mbar has been demonstrated yet with this technique.

Besides the investigations presented in~\cite{Dugrain,kitching}, other relevant studies on the optimized operation of compact 
ultra--high--vacuum (UHV) systems have been reported before. In~\cite{campo} the authors present a detailed analysis on the 
use of light--induced atomic desorption to modulate the background pressure of $^{87}$Rb atoms in a glass cell. They developed 
a model to find the number of atoms loaded in a MOT when the light is on, and demonstrated an order of magnitude increase under 
this condition. In the context of atom interferometry and atom sensors, an UHV system was designed and tested for operation in the
highly vibrating environment of a rocket~\cite{grosse}. In Ref.~\cite{scherer} the authors investigated the use of passive
vacuum pumps (non-evaporable getter pumps) for the development of compact cold atom sensors. Finally, in Ref.~\cite{basu} 
microfabricated non magnetic ion pumps were demonstrated with the aim of maintaining UHV conditions in miniature vacuum chambers 
for atom interferometry.

The aim of the present work is to understand from the physics point of view, the pressure dynamics of single vacuum chambers 
loaded with atoms via a dispenser and pumped out by a SIP. So far, SIPs are commonly used in cold atom experiments 
requiring UHV. Since they are an unavoidable component, which is at the same time able to provide
pressure readings~\cite{Audi,Coppa}, it is therefore relevant to have a physical model of the observed vacuum dynamics. This 
dynamics is not only determined by the pumping mechanism of the SIPs but also by conductance of the whole system 
and the dispenser sourcing effect. Understanding this dynamics would allow, for instance, the design of miniature 
SIPs~\cite{basu} and avoid the use of pressure gauges improving the compactness of the experiments.

To reach a good fidelity in estimating the pressure at the vacuum chamber, we develop an accurate calibration procedure to 
quantify the leakage ion current in the SIP. To achieve this goal, we first model the conductance of the vacuum system. Then, 
using the model and a protocol based on a pulsed dispenser current, we measure the temporal evolution of the pressure in the 
system. As we will see, the physical parameters describing the pressure dynamics extracted in this way, allow the reduction 
of the dead time in cold atom experiments by combining a fast loading rate of cold atom clouds (high partial $^{87}$Rb 
pressure regime) and a fast removal of background atoms after the production of these clouds~\cite{Dugrain}. It is worth 
mentioning that the commonly used models~\cite{SUETSUGU1995105, Ho1982} for the SIP pumping speed do not explain the important 
pressure variations (more than two orders of magnitude) we observed. In fact, on measurement time scales of several minutes 
the nature of the dominant pumping mechanism changes, and this effect needs to be taken into account.

\section{Steady state pumping behavior}
\begin{figure}[htb]
 \includegraphics[width=0.8\linewidth]{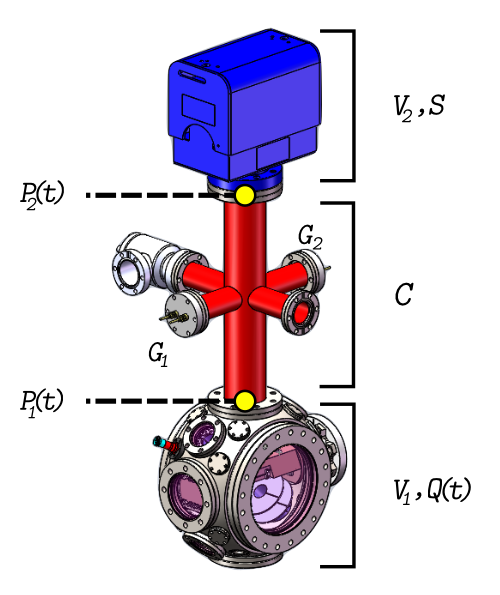}
\caption{Sketch of the considered experimental setup. The vacuum chamber of volume $V_1$ contains the atom source producing
a flow $Q(t)$. This gas at a pressure $P_1(t)$ in the chamber produces a pressure $P_2(t)$ at the pump through a pipe of 
conductance $C$. In the pump volume $V_2$, the atoms are pumped at a nominal speed $S$. Also represented in this figure are 
deactivated getter pumps $G_1$ and $G_2$.}
\label{singlechamber}
\end{figure}

In this manuscript we will consider a single vacuum chamber system as represented in Fig.~\ref{singlechamber}. It is a 
simplified configuration containing a chamber of volume $V_1$ with the atom source (dispenser) that produces a flow $Q(t)$ 
that goes to a pump with a 
nominal pumping speed $S$. The pump and the chamber are connected through a pipe with a conductance $C$. With these 
definitions, we can then relate the pressure at the chamber $P_1(t)$ to the pressure at the pump $P_2(t)$. In a steady 
state, neglecting leaks and in the free molecular regime, these quantities are related by the equation of the steady--state 
sourcing flux $Q(\infty)$
\begin{equation}
 Q(\infty) = C[P_{1}(\infty)-P_{2}(\infty)] = S P_{2}(\infty) = S_{\rm eff} P_{1}(\infty)\ ,
\label{eq:flow}
\end{equation}
where $S_{\rm eff}$ is the effective pumping speed seen by the chamber as determined by $C$. More generally, 
$P_{1}$ follows the gas balance equation~\cite{Chiggiato}
\begin{equation}
 V_1 \frac{dP_{1}(t)}{dt} = Q(t) - S_{\rm eff}P_{1}(t)\ .
\label{eq:gasbalance}
\end{equation}
Since the characteristic pumping time $\tau \equiv V_1/S_{\rm eff}$ controls the pressure transients in the vacuum system, 
$S_{\rm eff}$ needs to be properly determined. This is an important question in particular for cold atom experiments with 
time dependent sources of alkali atoms.

\subsection{Leakage current}
Normally, the pressure is translated into current readings by the pump controller. However, in the presence of alkali gases 
there exists a modification of the pump leakage current $I_{\ell}$~\cite{gammavac}. This modification is 
responsible for an overestimation of the real pressure. It originates from a thin layer of alkali ions stuck to the pump 
walls. Together with $I_{\ell}$ there is also an ion current $I$ produced by the ionization of the gas flowing through the pump 
electrodes. These two currents contribute to the measured current $I_{\rm m}$ actually reported by the pump controller (the 
current reading).

The leakage current $I_{\ell}$ is typically on the order of \mbox{100 nA} and it is usually neglected in high vacuum regimes 
(it corresponds to an overestimation of $\simeq$10$^{-9}$ mbar). However, neglecting this current affects the use of the 
ion pump  as a pressure gauge in the UHV regime~\cite{Chiggiato} ($<10^{-9}$ mbar). So, we will include the effect of 
this current in the analysis below.

Following ion pump manufacturers and taking into account that $I=I(P_2)$ is a function of the pressure at the pump, we have 
the following expression for $I_{\rm m}$
\begin{equation}
 I_{\rm m} (U) = I(P_2) + I_{\ell} = f(P_2)\cdot U + I_{\ell}\ ,
\label{eq:current}
\end{equation}
where $U$ is the applied voltage between the pump electrodes. From this equation we see that an accurate determination of the 
actual pressure ($P_2$ or $I$) requires a precise knowledge of the leakage current. The usual way of finding $I_{\ell}$ is 
to measure $I_{\rm m}$ while the pump's magnets are removed. In this situation, there is no ionization process and we have 
$I=0$ A. However, this method requires the pump to be stopped and does not allow a real--time monitoring of the pressure. 

\begin{figure}[htb]
 \includegraphics[width=0.9\linewidth]{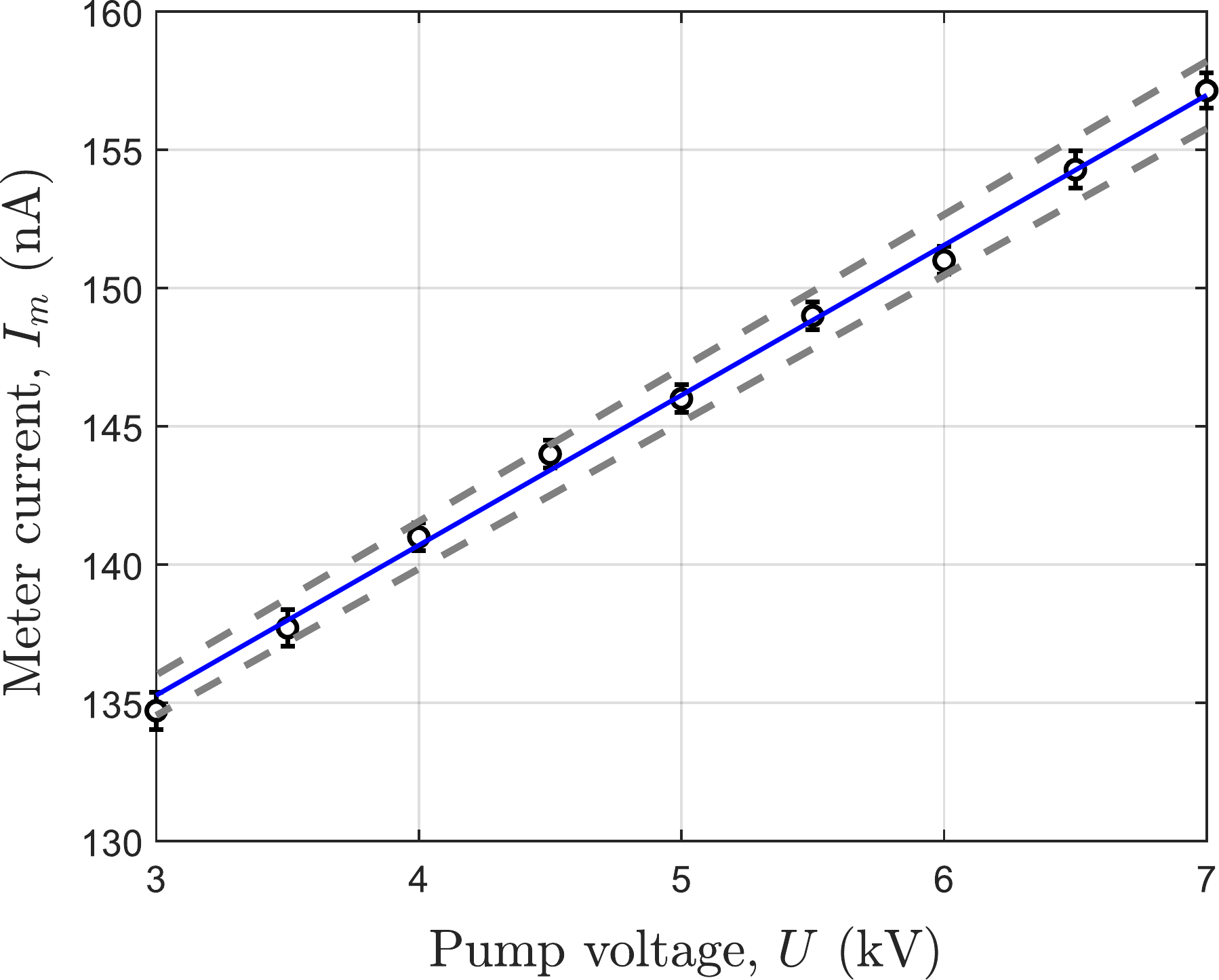}
\caption {Current-voltage (I--V) characteristic of the pump. The leakage current $I_{\ell}$ = 119.0 $\pm$ 0.4 nA is obtained 
from a linear fit (solid line) of the measured current $I_{\rm m}$. Dashed lines represent the confidence interval of the fitting 
parameters.}
\label{fig:LeakageCurrent}
\end{figure}

Here, we measure $I_{\ell}$ by gradually decreasing $U$ within the nominal working range of the pump~\cite{gammavac}. Following  
Eq.~(\ref{eq:current}), the leakage current is then determined by extrapolating the data to $U=0$ V. The result of this 
measurement is presented in Fig.~\ref{fig:LeakageCurrent}, where the observed linear behavior indicates that the pressure 
$P_2$ does not depend on the applied voltage $U$ at the pressure levels we performed the experiment.
The obtained value of the leakage current is \mbox{$I_{\ell} = 119.0 \pm 0.4$ nA}. As we will see in the following section, 
the accuracy in the pressure measurement obtained with this method allows us to model the pumping dynamics for pressures 
$<10^{-9}$ mbar.

\subsection{Determination of the pressure at the chamber from the SIP current}
Once the leakage current is found, we can evaluate the ion current inside the pump $I(P_2)$ using Eq.~(\ref{eq:current}) and 
the current reading $I_{\rm m}$. The next problem is then to determine the explicit dependence of the ion current on the 
pressure at the pump, $f(P_2)$. Then we can invert the function $f(P_2)$ and, in the steady state regime, compute the pressure 
at the chamber using Eq.~(\ref{eq:flow}), namely 
\begin{equation}
 P_{1} = \left(\frac{S}{C} + 1\right)P_{2}\ .
\label{eq:pressuregradient}
\end{equation}

In the free molecular regime, the conductance $C$ depends only on the geometry of the vacuum system for a given gas 
species and temperature. Using the Santeler equation~\cite{Santeler} for the transmission 
probability through a 
cylindrical pipe of radius $R$ and length $L$, we can calculate the conductance for a molecule of mass $m$ at room 
temperature using the relation~\cite{Chiggiato}
\begin{equation}
 C = 11.75 \pi R^{2} \sqrt{\frac{m_{\rm N_{2}}}{m}} \left[1+\frac{3L}{8R}\left(1+\frac{1}{3(1+L/7R)}\right)
 \right]^{-1} \ .
\label{eq:conductance}
\end{equation}
In Eq.~(\ref{eq:conductance}) $m_{\rm N_{2}}$ is the mass of a nitrogen molecule, and $R$ and $L$ must be expressed 
in cm to get $C$ in L.s$^{-1}$. Now, let's get an estimate of the value of $C$ for our vacuum system. In a constant flow 
regime, the conductance of our particular geometry (central pipe of $L$ = 35.2 cm and $R$ = 3 cm) evaluates to 
$C =$ 32 L.s$^{-1}$ for the $^{87}\text{Rb}$ monoatomic gas. This value is obtained neglecting contributions from 
the cross which is a reasonable assumption in the constant flow regime. Finding $S$ precisely is slightly more difficult when 
considering pumping of $^{87}$Rb atoms. However, following the pump's manufacturer documentation~\cite{gammavac} we can use the 
linear relation
\begin{equation}
 P_2=\alpha k \frac{I}{U} \ ,
\label{eq:ManfacturerConversion}
\end{equation}
to express the pressure at the pump in terms of the ion current. Here, $k$=10.9 mbar.V.A$^{-1}$ at room temperature 
and $\alpha$ is a calibration factor. This factor is the ionization vacuum gages' correction factor which links  
the pressure measurement of specific gas species to calibration measurements using nitrogen. For Rb 
$\alpha=4.3$~\cite{summers}. As we will see later, the empirical relation Eq.~(\ref{eq:ManfacturerConversion}) does not take 
into account the fact that the actual relationship between the pressure $P_2$ and the ion current in the pump $I$ is nonlinear.

\begin{figure}[htb]
\includegraphics[width=\linewidth]{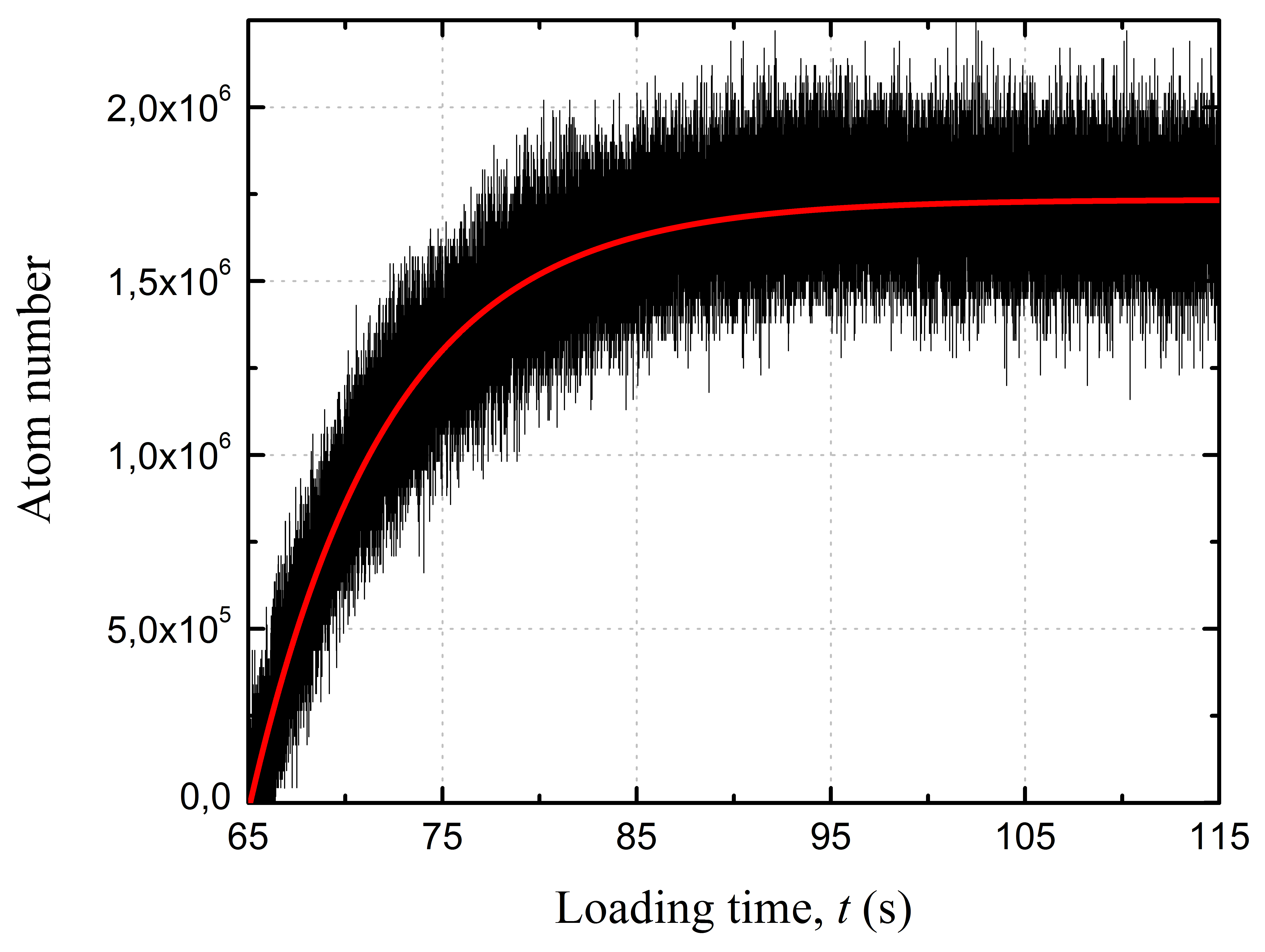}
\caption{Number of $^{87}$Rb atoms loaded in the MOT (black) for a dispenser current of 4.75 A. The fit (red solid line) to 
the experimental data gives a characteristic loading time of $7.13 \pm 0.02 ~\text{s}$.}
\label{fig:MOTloading}
\end{figure}

With our MOT, we can realize an independent measurement of $P_1$ instead of 
computing it using  Eq.~(\ref{eq:pressuregradient}). From the loading curve of the MOT, as shown in 
Fig.~\ref{fig:MOTloading}, we can find $P_1$ as indicated in~\cite{Sackett,moore}. In 
fact, the loading dynamics of the MOT critically depends on the background pressure of the trapped species and other gases. As 
has been 
demonstrated in the past~\cite{Monroe, Dugrain, Sackett,moore}, these curves produce reliable pressure measurements. In our 
experimental setup we use a mirror MOT obtained with an atom chip. The relevant experimental details are as follows:  the 
red-detuned cooling lasers (-1.5 $\Gamma$ where $\Gamma=2\pi\times$6 MHz is the natural line width of $^{87}$Rb $D_{2}$ line)  
have a maximum power of $\simeq$40 mW shared by four independent MOT beams of about 2.5 cm of 1/$e^2$ diameter. The 
magnetic field gradient is 11 G.cm$^{-1}$. During 100 s of loading, the fluorescence emitted by the atoms is collected on a 
photodiode with a solid angle of 1.3$\times 10^{-2}$ srad. This signal is used to compute the atom number. In order to vary 
the pressure $P_1$, we change the dispenser current to produce different stationary gas flows $Q$. 

\begin{figure}[htb]
\includegraphics[width=\linewidth]{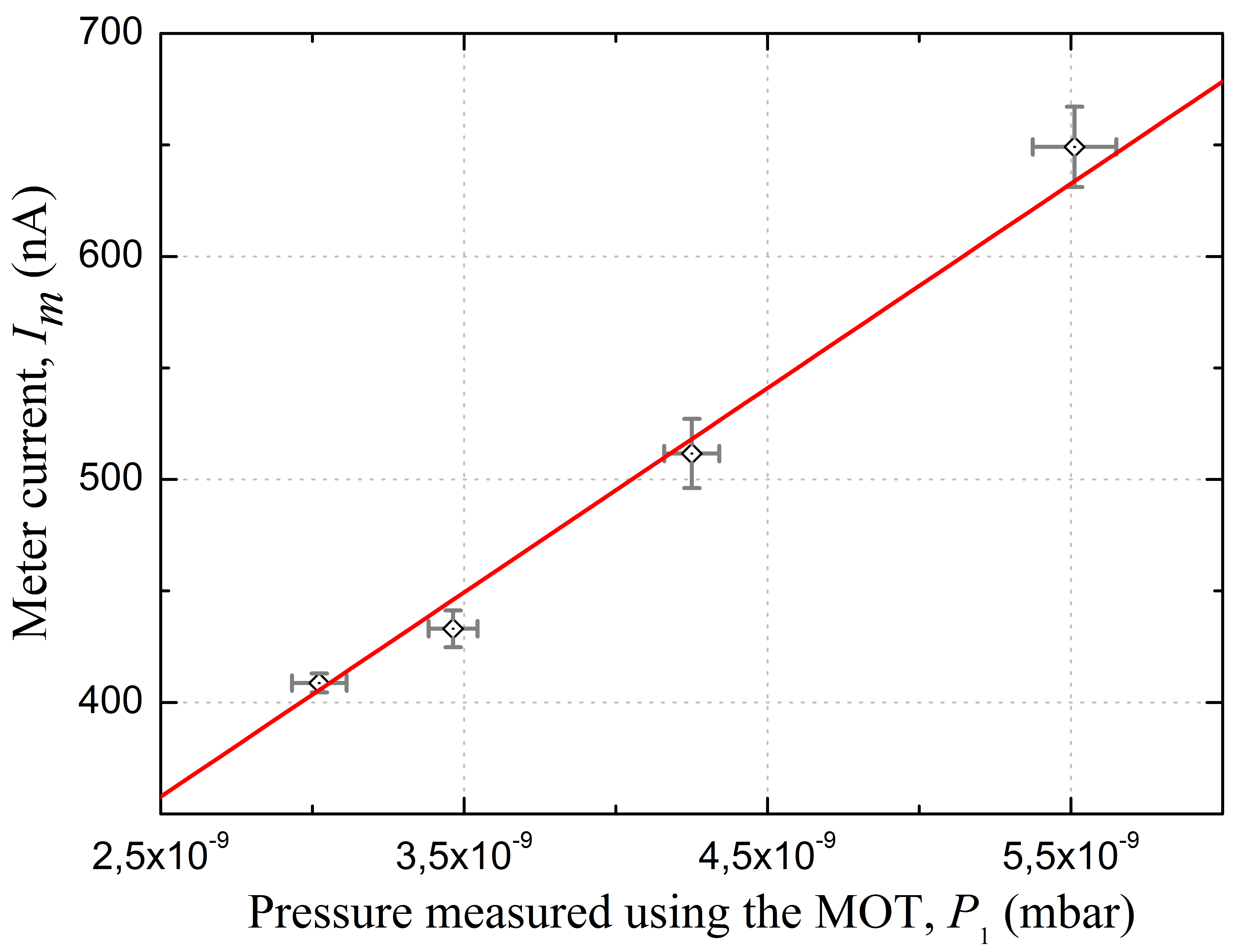}
\caption{Meter current vs.\ measured pressure $P_1$ from MOT loading curves (points). The leakage current $I_{\ell}$ and the 
parameter $\beta$ extracted from the fit (solid lined), using Eq.~(\ref{eq:currenttopressurefit}), are respectively equal to 
\mbox{130 $\pm$ 20 nA} and \mbox{(9.2 $\pm$ 0.6)$\times 10^{10}$ nA.mbar$^{-1}$}.}
\label{fig:pressureresult}
\end{figure}

In analogy to Eq.~(\ref{eq:ManfacturerConversion}), we assume that the ion current $I$ is proportional to the pressure at 
the vacuum chamber $P_1$, measured with the MOT. That is $I=\beta P_1$, where $\beta$ is a parameter to be experimentally 
determined. Then, we can write the following equation for the ion current reading as a function of the pressure
$I_{\rm m}$ 
\begin{equation}
 I_{\rm m} = I_{\ell} + \beta P_1 \ .
\label{eq:currenttopressurefit}
\end{equation}
In Fig.~\ref{fig:pressureresult} we plot the dependence of $I_{\rm m}$ on the pressure $P_1$ measured from MOT loading curves 
at steady state. This result offers one independent method to validate the assumption leading to 
Eq.~(\ref{eq:currenttopressurefit}). This method consists in finding the leakage current from MOT measurements and comparing 
the obtained value with the one extracted from the I--V characterization. Fitting the data in Fig.~\ref{fig:pressureresult}
using Eq.~(\ref{eq:currenttopressurefit}), we find for $I_{\ell}$ a 
value of \mbox{130 $\pm$ 20 nA}, in good agreement with the result given by the I--V characterization presented in 
Fig.~\ref{fig:LeakageCurrent}. This agreement supports the use of $\beta$ to compute the pressure in the vacuum chamber by the 
relation $P_1=I/\beta$. For the parameter $\beta$ we obtain the value of \mbox{(9.2 $\pm$ 0.6)$\times 10^{10}$ nA.mbar$^{-1}$}.

It would be tempting to use the information from equation (\ref{eq:pressuregradient}) and Eqs.~(\ref{eq:ManfacturerConversion}) 
and (\ref{eq:currenttopressurefit}) to find the pumping speed $S$. However, as we will see in the next section, constant 
pumping speeds do not properly describe the transient behavior of the pressure when switching on and off the dispenser current.

\section{Analysis of the nonlinear pumping dynamics }
\subsection{Derivation of the dynamics}
When searching for the reduction of the vacuum system contribution to the dead time between interferometric measurements, we 
need to focus on the pump--down dynamics that is triggered after loading the MOT and switching off the atom source 
(dispenser). To achieve this goal, we devised a pressure measurement protocol which is as follows: first, we switch on 
the dispenser at a given current and monitor the pressure rise until 
it reaches the steady state~\cite{moore}. The current ranges from 3.75 A to 5 A, with a step of 0.25 A.  Then, we switch off 
the dispenser 
and record the pressure decay (pump--down curve) until it goes back to the steady state. We allow both of the transient 
processes to last for about $\simeq$1000 s.

\begin{figure*}[htb]
 \includegraphics[width=0.5\linewidth]{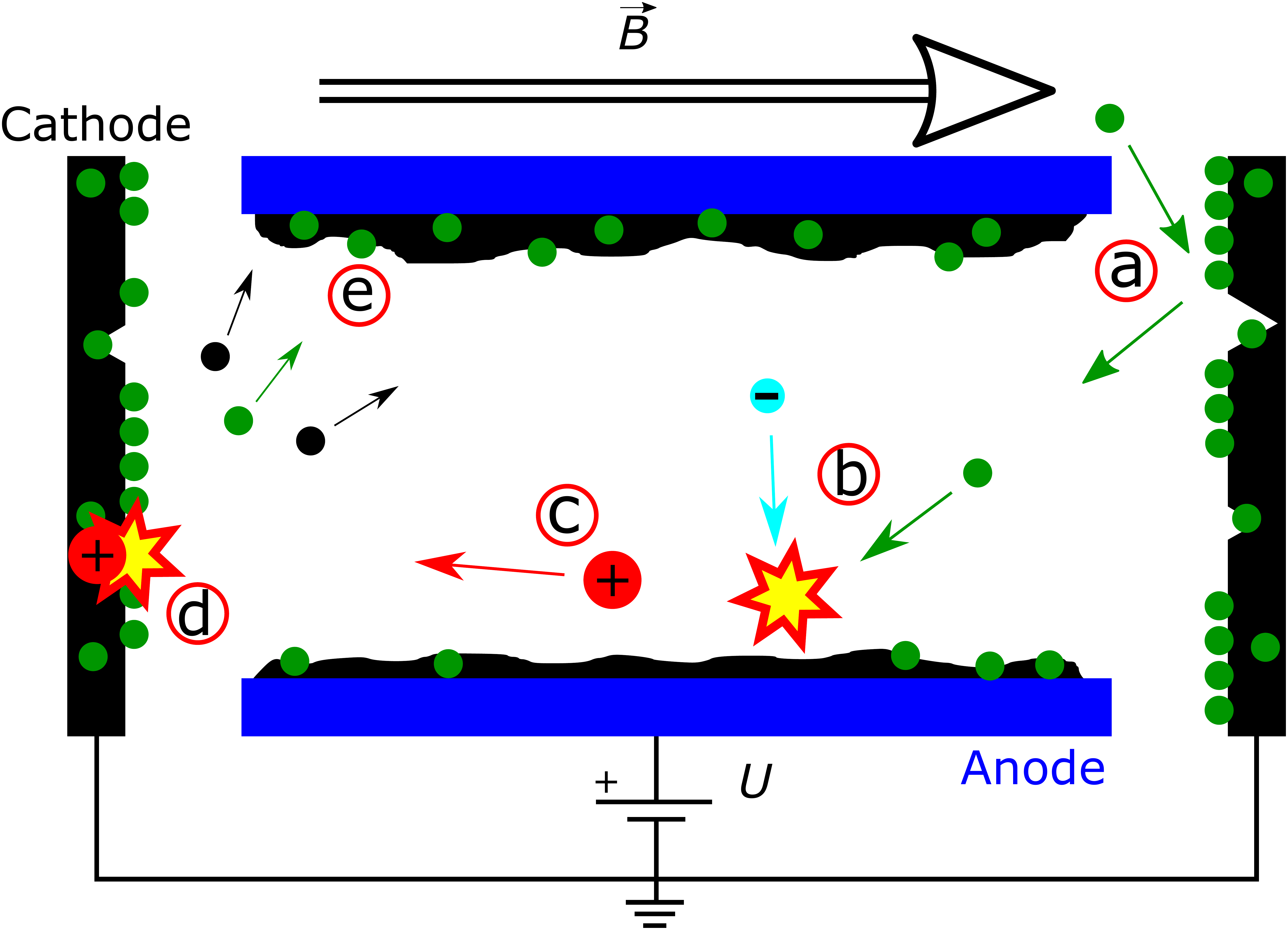}
 \caption{Representation of the pumping unit cell (of Penning type). A high voltage $U$ is applied between the anode (blue),
covered by some getter material (black), and the cathode (black). A particle in the gas (green) enters the pumping cell and
hits the cathode (a) where it is stuck or deflected towards the anode. On its way to this electrode, the particle
collides with an electron (b) and gets ionized. The ion is then accelerated towards the cathode (c) with
eventually enough energy to be buried and sputter pumping material (d). The freshly sputtered material covers the
internal walls of the cylindrical anode (e) which in then ready to pump more particles.}
 \label{fig:SIP}
\end{figure*}

In the following we will develop a mathematical formalism to describe the main physical processes taking place during the 
pump--down dynamics. Firstly, we will assume that the dispenser is no longer sourcing atoms into the chamber after being 
switched off. In this case, we can consider that the pressure evolution is mainly due to the pumping by the SIP in the 
presence of a residual outgassing flow $Q(t)$ coming from the vacuum chamber. In steady state $Q(t)$ will be solely given 
by the thermal outgassing in the system. Secondly, we will suppose that the pump contains an ensemble of Penning cells with 
the geometry sketched in Fig.~\ref{fig:SIP}. Thirdly, let's assume that at the time instant $t$:
\begin{enumerate}
 \item there already exists some sputtered pumping material (e.g.\ Ti) that pumps the gas, reducing the pressure by an 
 amount $-a P_2(t) dt$ [process \circled{a} in Fig.~\ref{fig:SIP}];
 \item some trapped molecules are released by the incident ion flux increasing the pressure by $c I(t) dt$ 
 [process \circled{d} in Fig.~\ref{fig:SIP}];
 \item some pumping material sputtered by the ion flux pumps the gas, reducing the pressure by $-a P_2(t) \times b I(t) dt$ 
 [process \circled{e} in Fig.~\ref{fig:SIP}].
\end{enumerate}

In point 1 above, the coefficient $a$ represents the probability rate at which a particle reaching the cathode 
(made out of a pumping material) sticks to it. Furthermore, when the gas molecules gets ionized inside the pump [\circled{b} in 
Fig.~\ref{fig:SIP}], the 
applied voltage accelerates the ions [\circled{c} in Fig.~\ref{fig:SIP}] towards the cathode. If the ions have sufficient 
energy they can release previously trapped particles (when they collide with the walls) with a desorption rate proportional to 
$c$ (point 2) and also, they can sputter pumping material with a yield characterized by the coefficient $b$ (point 3).

The physical processes we just described are in agreement with the fact that the pumping speed of the SIP decreases when the 
pressure decreases. The reason is the decrease of the discharge intensity (current per unit pressure) in this situation. This 
reduction of the pumping speed depends strongly on the pump parameters such as the applied anode voltage, the 
magnetic field, and the geometry of the pumping cell.

Collecting together the above mentioned processes, we arrive at the following differential equation for the pressure 
evolution at the pump
\begin{equation}
 \frac{dP_2(t)}{dt} = -a P_2(t) -a P_2(t)\times b I(t) + c I(t) + \frac{Q(t)}{V_2}\ ,
\label{eq:pressure1}
\end{equation}
where $V_2$ is the pump volume. In the next section we use this model to fit the experimental data and determine 
the physical parameters defining the nonlinear dynamics.

\subsection{Practical fitting model}
Instead of working directly with the pressure equation~(\ref{eq:pressure1}), here we will derive a practical model that will 
allow 
a fitting of the experimental data. Our meter outputs current values and therefore, it would be more natural to work 
with the ion current $I(t)$ rather than the pressure $P_2(t)$. However, the physical processes we just discussed indicate that 
we cannot use Eq.~(\ref{eq:ManfacturerConversion}) to relate these quantities. Indeed, $I(t)$ has a nontrivial dependence on 
the pressure governed by the pressure regime the pump is working in. This fact is encoded by the empirical equation~\cite{Audi}
\begin{equation}
 I(t) = h P_2(t)^n \ ,
\label{eq:NLioncurrent}
\end{equation}
where the exponent $n$ is a real number used to identify the different pressure regimes. It depends on the gas species and 
the geometry of the pump, and will be determined from the fitting procedure. In 
Eq.~(\ref{eq:NLioncurrent}), $h$ is a time independent calibration parameter defined by the type and size of the 
pump.

\begin{figure}[htb]
\subcapraggedrighttrue
\subfigure[\ Dispenser currents: 5 A (black circles), 4.75 A (blue diamonds) and 4.5 A (red squares).]
 {\includegraphics[width=\linewidth]{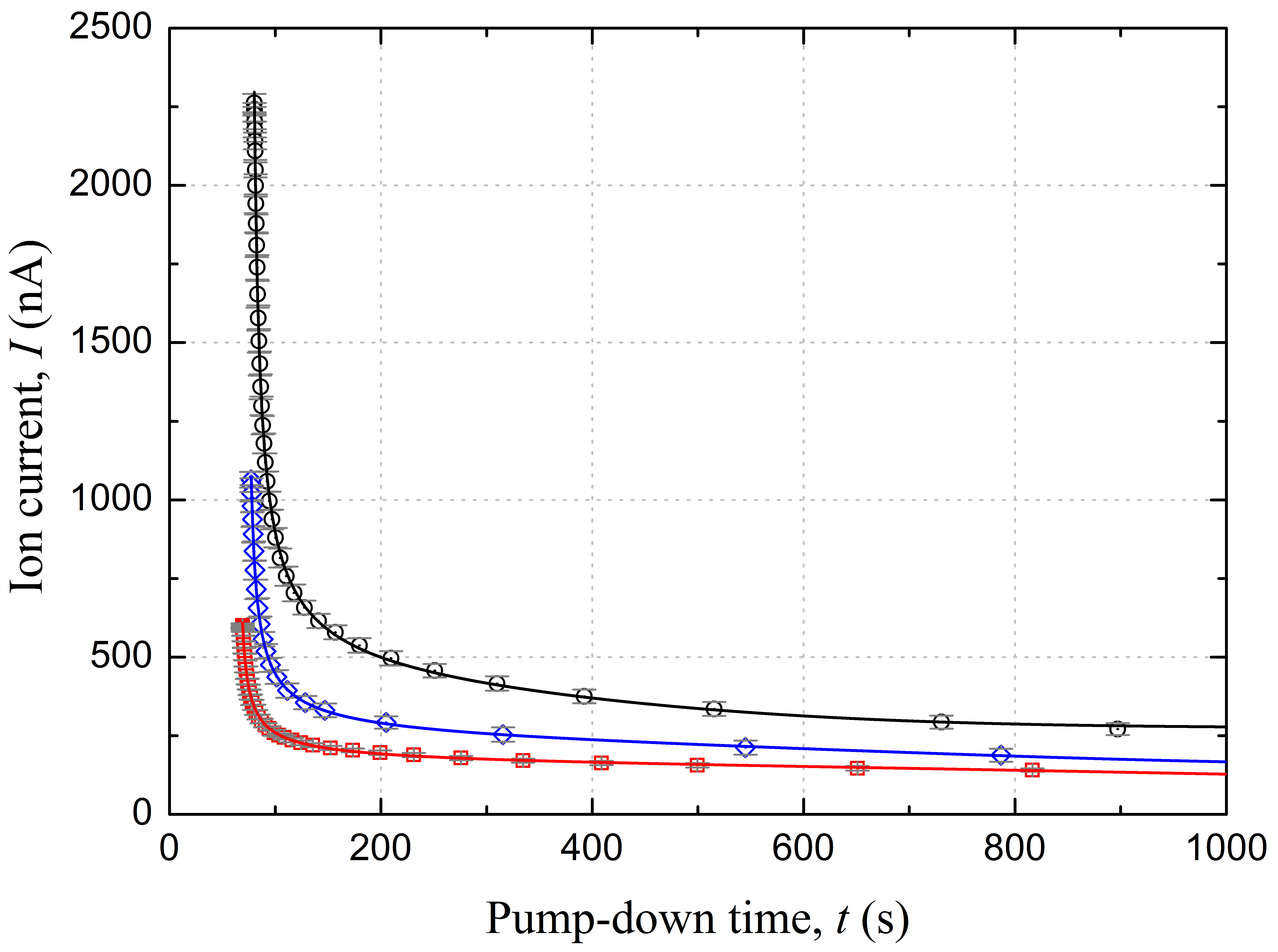}}
\subfigure[\ Dispenser currents: 4.25 A (black circles), 4 A (blue diamonds) and 3.75 A (red squares).]
 {\includegraphics[width=\linewidth]{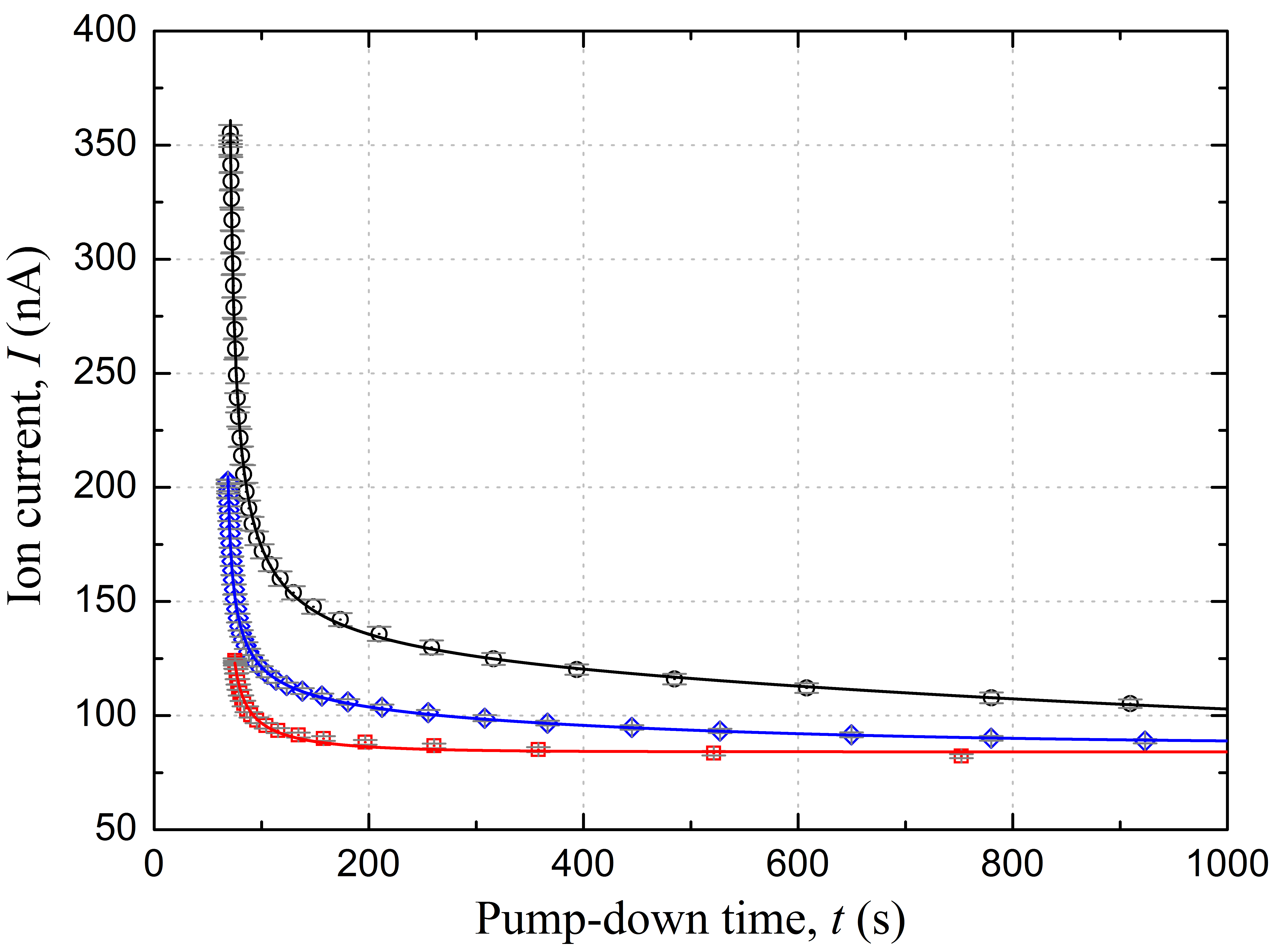}}
\caption{We record the pump--down curves (data points) after switching off the dispenser currents, initially at levels 
given in captions (a) and (b). In each case, the measured ion current $I(t)$ is fitted (solid lines) by numerical 
integration of Eq.~(\ref{eq:FittedEquation}).}
\label{fig:Fits}
\end{figure}

Inserting Eq.~(\ref{eq:NLioncurrent}) into (\ref{eq:pressure1}) we obtain the following equation in terms 
of the ion current
\begin{equation}
 \frac{dI(t)}{dt} = -\alpha_1 I(t) - \alpha_2 I(t)^2 + \left[ \alpha_3 I(t) + q \right] I(t)^{1-\frac{1}{n}}\ ,
\label{eq:FittedEquation}
\end{equation}
with $\alpha_1 \equiv na$, $\alpha_2 \equiv n a b$, $\alpha_3 \equiv n c \sqrt[n]{h}$, 
$q \equiv n \sqrt[n]{h} Q_{\rm th}/V_2$. These parameters will be treated as independent and used in the fitting procedure. When 
writing (\ref{eq:FittedEquation}) we considered 
that after switching off the dispenser $Q(t)$ reaches the constant thermal outgassing flux $Q_{\rm th}$ in a time scale shorter 
than the time frame required to reach the steady state. As will we see later, such an approximation is compatible with our 
observations. We measured the pump--down curves presented in Fig.~\ref{fig:Fits}. The points are the experimental data and 
the solid lines are fits obtained with Eq.~(\ref{eq:FittedEquation}). As can be seen in this figure, there is 
a very good agreement between the theory and the experimental data. 

\begin{figure*}[htb]
\subcapraggedrighttrue
 \subfigure[\ Pumping regime parameter.]
 {\includegraphics[width=0.45\linewidth]{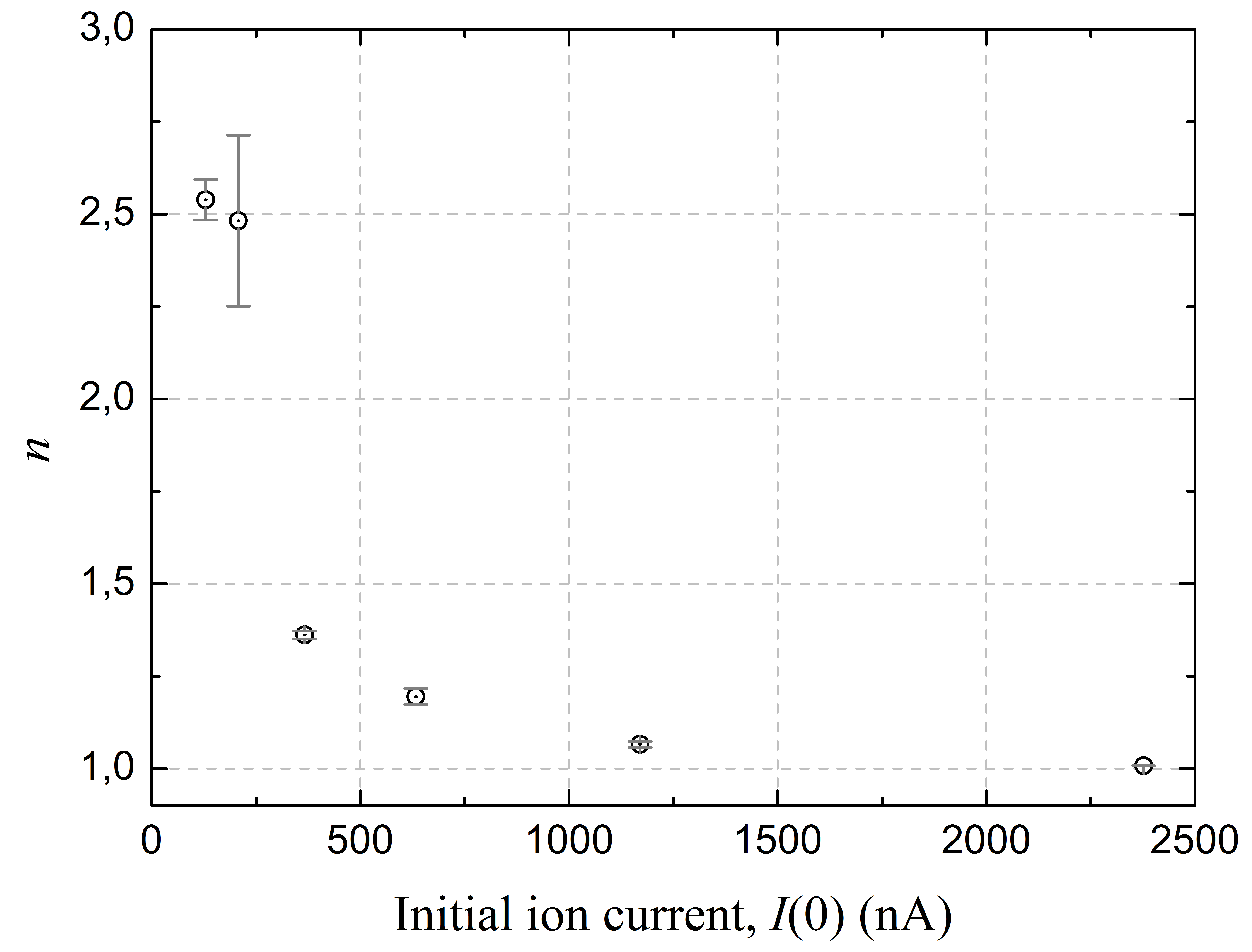}\label{fig:n}}
\subfigure[\ Former sputtered material pumping rate parameter.]
 {\includegraphics[width=0.45\linewidth]{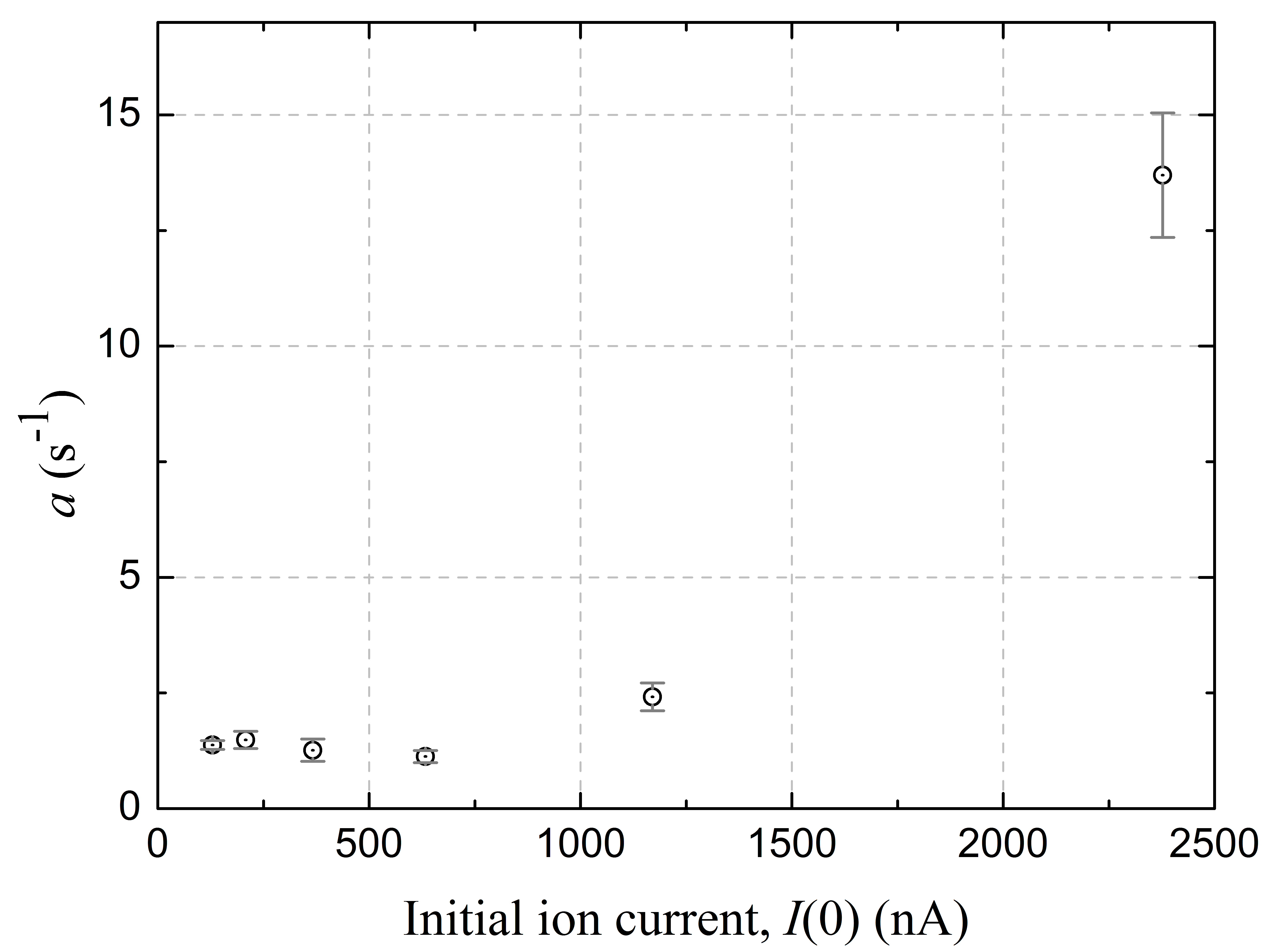}\label{fig:a}}
\subfigure[\ Instantaneous ion flux induced sputtering pumping rate parameter.]
 {\includegraphics[width=0.465\linewidth]{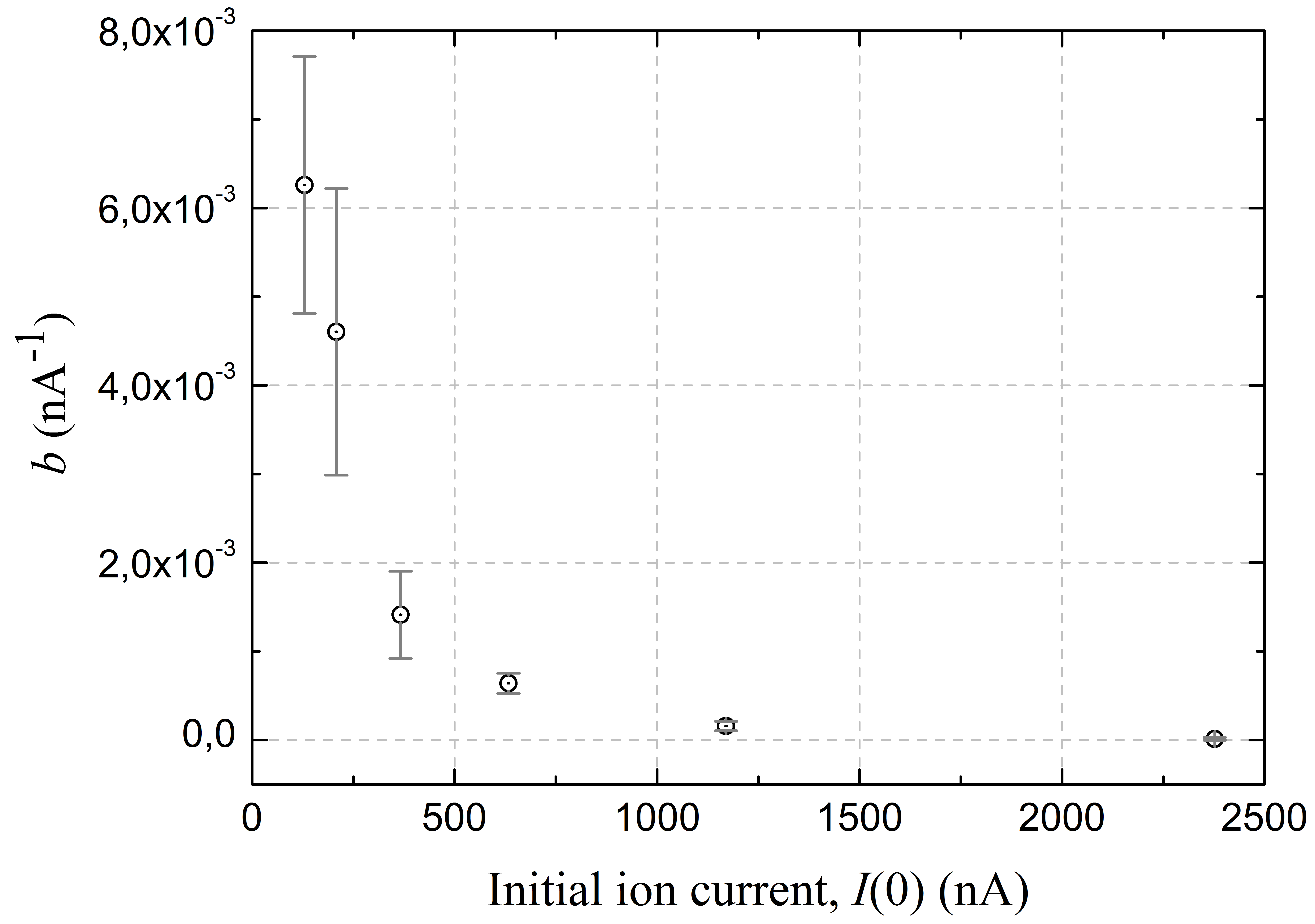}\label{fig:b}}
\subfigure[\ Ion flux desorption parameter.]
 {\includegraphics[width=0.435\linewidth]{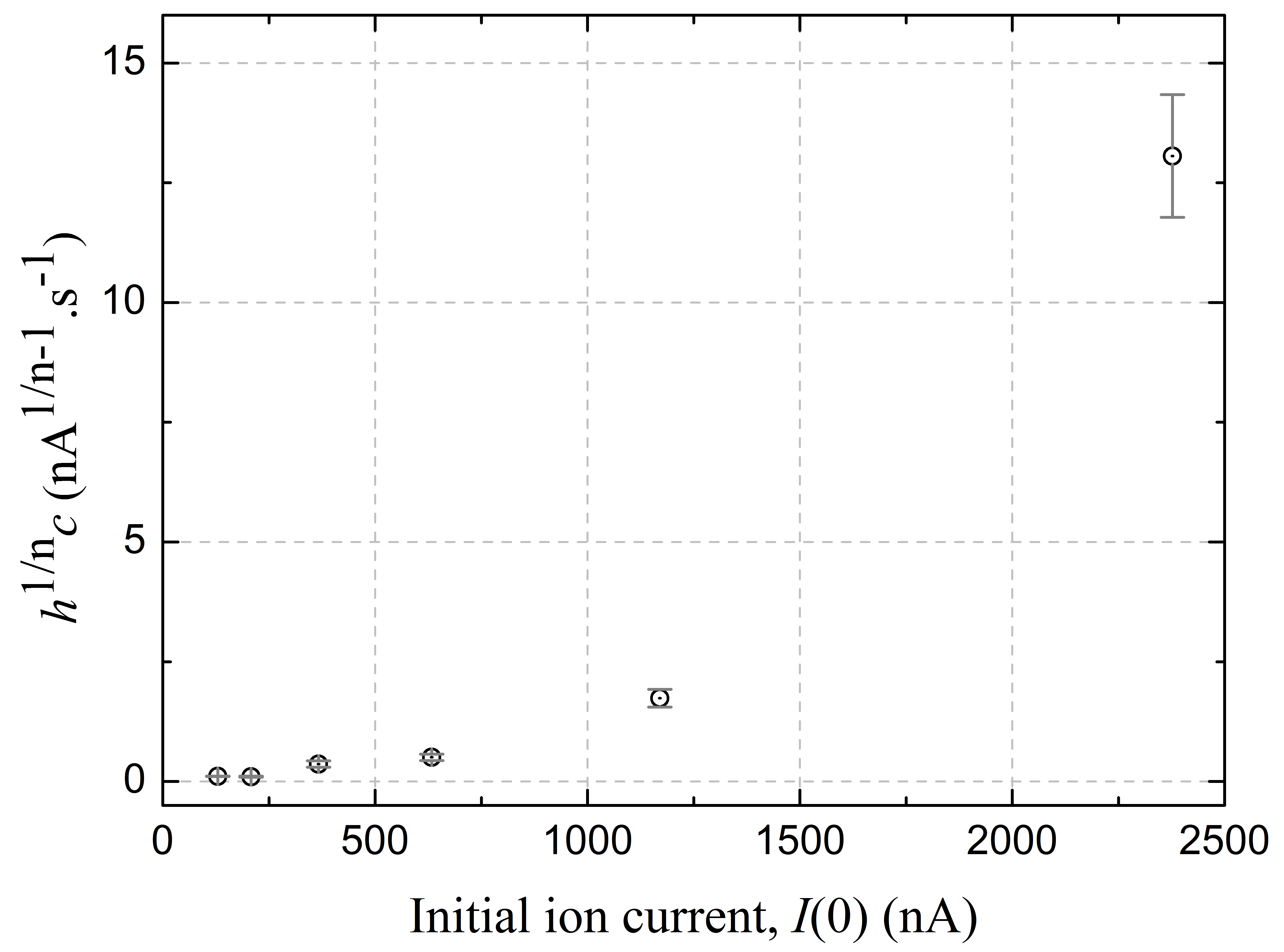}\label{fig:c}}
\subfigure[\ Thermal outgassing source parameter.]
{\includegraphics[width=0.45\linewidth]{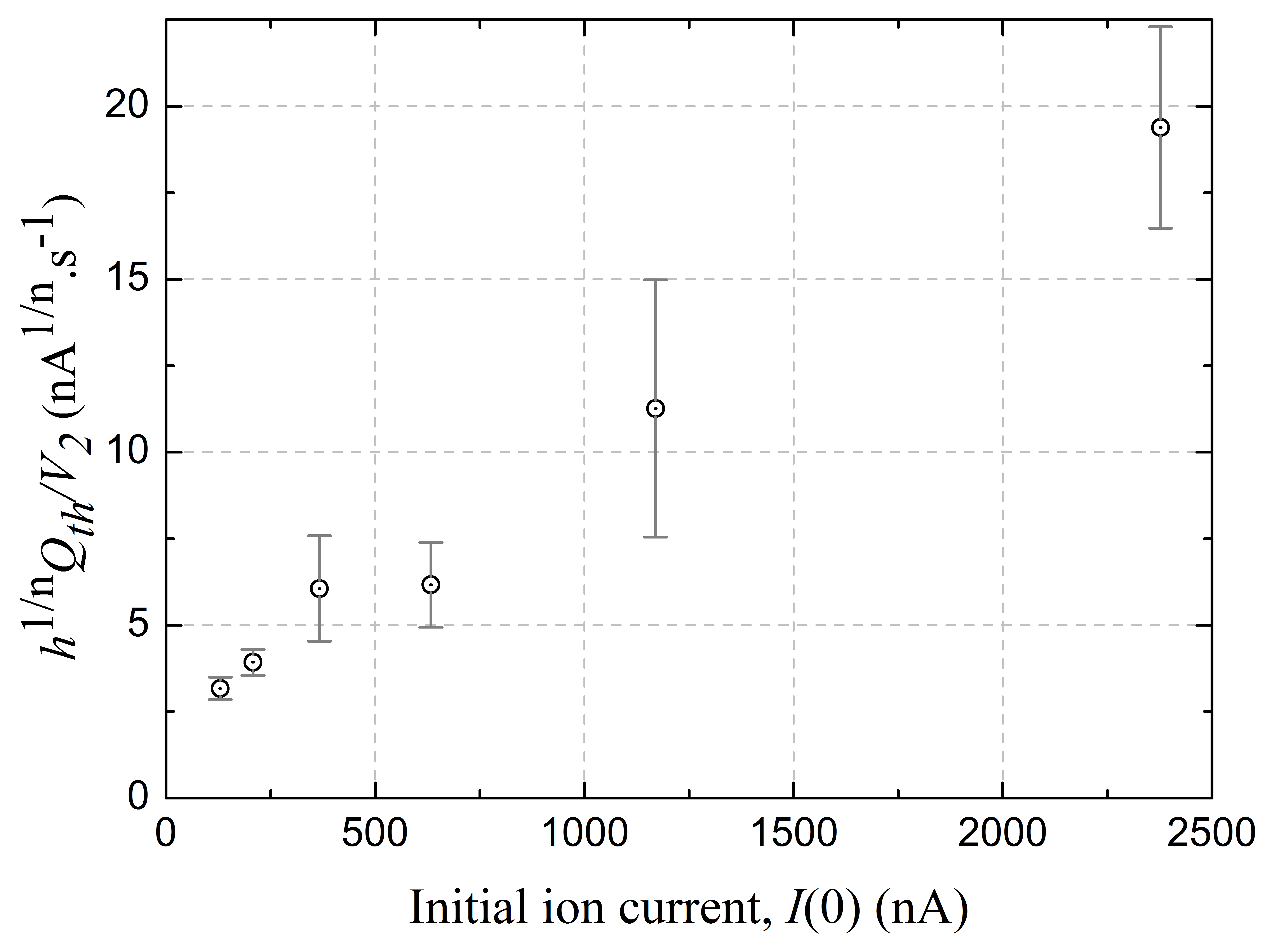}\label{fig:Q}}
\caption{Dependence of the parameters describing the physical processes in the pump volume on the initial ion current (just 
after turning off the dispenser current). These parameters result from the fitting of the pump--down curves in 
Fig.~\ref{fig:Fits} using the numerical solution of the differential equation~(\ref{eq:FittedEquation}).}
\label{fig:Prameters}
\end{figure*}

To validate the model beyond the criteria set by the fit 
quality, we study the dependence of the fitting parameters on the pressure $P_{2}$ looking at their behavior in different 
pressure regimes. The measurement protocol used is as follows: we change the dispenser current and wait until the pressure 
reaches an equilibrium state. Next, we measure the ion current at this equilibrium situation, before switching off the 
dispenser. Finally, we start the measurement of the pump--down dynamics. The results obtained with this protocol are presented
in Fig.~\ref{fig:Prameters}. It shows the dependence of the fitting parameters on the initial ion current.

As expected, the value of $n$ increases when the pressure goes down~\cite{Dallos1967} as can be seen in Fig.~\ref{fig:n}. 
Moreover, it reaches unity at the highest measured pressure. The obtained value in this later case is actually compatible with 
pump manufacturers' reported values for air. At low pressures it goes beyond 1.5, a value never reported before to our 
knowledge and that might be in agreement with the fact that we are pumping an alkali gas.

The parameter $a$, according to our model, depends on the pump's cathode geometry and the sticking factor, this later being 
a function of the temperature and the gas species. From the measurement in Fig.~\ref{fig:a} we see that at low initial 
ion currents (pressures) $a$ is 
relatively constant. This is expected since in this case the sticking probability 
should correspond to a linear process given the gas density in the pump. However, when the initial ion current is increased, $a$ 
eventually increases, suggesting a modification of the sticking probability. This is coherent with the fact that in this 
situation the behavior of $n$ also indicates a change in the pressure regime. We attributed the change of $a$ with the initial 
ion current to the change in the sticking probability because the cell geometry does not change.

From Fig.~\ref{fig:b} we see that the parameter $b$ tends to zero when the initial ion current is increased. This is also an 
expected behavior since this parameter is related to the discharge current which is depressed by the space--charge effect when 
the pressure rises. As a consequence, the sputtering rate becomes reduced~\cite{Audi}. In fact, what happens is that at 
relatively high initial ion currents or pressures, the energy of the ions hitting the cathode is no longer exclusively defined 
by the applied voltage $U$. 

In order to interpret the behavior of the parameters $c$ and $Q_{\rm th}$ we need to isolate them from the calibration factor 
$h$. This requires us to perform independent measurements. However, it is fair to consider $h$ as a scaling factor in 
Fig.~\ref{fig:c} and Fig.~\ref{fig:Q}. In this situation the increase of $c$ with the initial ion current might be a 
consequence of the bombardment boost in the presence of a significant number of gas particles in the pump volume. This process 
naturally leads 
to a relatively higher desorption rate of buried molecules. Increasing the initial ion current also leads to an increase in the 
thermal outgassing flux $Q_{\rm th}$ in the time scale we record the data ($\sim$1000 s). This effect is already observable in 
Fig.~\ref{fig:Fits} where the steady state ion current value depends on the dispenser current.

Finally, the pressure in the chamber can be determined using the coefficient $\beta$ obtained from the calibration measurement 
in Fig.~\ref{fig:pressureresult} and the expression $P_1(t) = I(t)/\beta$. This linear relation indicates that 
the science chamber, at pressure $P_1(t)$, acts as a {\it passive} particle reservoir feeding the ion current 
$I(t)$. On the contrary, the Eq.~(\ref{eq:NLioncurrent}) indicates that the pump volume, at a pressure $P_2(t)$, can be seen as 
an {\it active} particle reservoir feeding $I(t)$ because of the physical processes involving the gas inside the pump. 

Having these ideas in mind, we can outline the next calibration and monitoring protocols for the time evolution of 
the pressure in the vacuum chamber:
\begin{itemize}
 \item Calibration protocol
 \begin{enumerate}
  \item Use the MOT to perform a steady state calibration measurement as in Fig.~\ref{fig:pressureresult}.\label{it:i1}
  \item From the calibration step~\ref{it:i1} and Eq.~(\ref{eq:currenttopressurefit}) find $\beta$. \label{it:i2}
  \item Measure pump--down curves as in Fig.~\ref{fig:Fits} for different initial ion currents $I(0)$.
  \item Using Eq.~(\ref{eq:FittedEquation}), fit the pump--down curves to find the ion current $I(t)$.\label{it:i3}
  \item Find the physical parameters of this model for the different $I(0)$ and fit the data in Fig.~\ref{fig:Prameters}.
 \end{enumerate} 
 \item Monitoring protocol 
  \begin{enumerate}
  \item In a given experimental run, and before launching the relevant measurement, find $I(0)$.
  \item Compute the corresponding physical parameters from the fits done on Fig.~\ref{fig:Prameters}. 
  \item Use $\beta$ from step~\ref{it:i2} and the solution $I(t)$ from step~\ref{it:i3} to compute the pressure in the 
    chamber as $P_1(t) = I(t)/\beta$.
 \end{enumerate}
\end{itemize}

\section{Conclusion}
In this work we have developed a detailed physical model of the nonlinear pressure dynamics in sputter ion pumps. This 
model has been experimentally corroborated by the measured data. It includes parameters describing the complex physical 
processes taking place inside the vacuum pump. Moreover, we characterized the system conductance and used 
pressure measurements with a MOT to establish a link between the pressure in the vacuum chamber and the ion current 
provided by the pump. From a practical point of view, this relationship allows the pump current to be used as a good 
indicator of the pressure in the science chamber. From the observed dynamics, we can tailor the effective pumping 
speed and optimize the MOT loading time with respect to the contradictory requirements of having high repetition rates 
and high number of atoms in a single chamber. We hope that the physics investigated in this work will be useful 
in the future to engineer miniature and microscopic scale ion pumps~\cite{basu} for cold atom based compact quantum sensors.

\section{Acknowledgment}
This work has been funded  by the D\'el\'egation G\'en\'erale de l'Armement (DGA)
through the ANR ASTRID program (contrat ANR-13-ASTR-0031-01), by the UK Defence Science and Technology
Laboratory, under Grant No.\ DSTLX-1000097814 (DSTL--DGA PhD fellowship program), the Institut Francilien de Recherche 
sur les Atomes Froids (IFRAF), and the Emergence-UPMC program (contrat A1-MC-JC-2011/220). BMG acknowledges the support 
of the UK Quantum Technology Hub for Sensors and Metrology (UK Engineering and Physical Sciences Research Council 
grant EP/M013294/1).


\end{document}